\documentclass[pdflatex,sn-mathphys-num]{sn-jnl}

\usepackage{float}

\usepackage{graphicx}
\usepackage{multirow}
\usepackage{amsmath,amssymb,amsfonts}
\usepackage{amsthm}
\usepackage{mathrsfs}
\usepackage[title]{appendix}
\usepackage{xcolor}
\usepackage{textcomp}
\usepackage{manyfoot}
\usepackage{booktabs}
\usepackage{algorithm}
\usepackage{algorithmicx}
\usepackage{algpseudocode}
\usepackage{listings}

\theoremstyle{thmstyleone}

\theoremstyle{thmstyletwo}

\theoremstyle{thmstylethree}

\begin{document}

\title[Article Title]{\begin{center}
Terrain-aware Deep Learning for Wind Energy Applications: From Kilometer-scale Forecasts to Fine Wind Fields
\end{center}}

\author[1]{\fnm{Chensen} \sur{Lin}}
\email{linchensen@fudan.edu.cn}
\equalcont{These authors contributed equally to this work.}

\author[1]{\fnm{Ruian} \sur{Tie}}
\email{tieruian@fudan.edu.cn}
\equalcont{These authors contributed equally to this work.}

\author[1]{\fnm{Shihong} \sur{Yi}}
\email{yishihong@fudan.edu.cn}

\author[1]{\fnm{Xiaohui} \sur{Zhong}}
\email{zhongxiaohui@fudan.edu.cn}

\author*[1]{\fnm{Hao} \sur{Li}}
\email{lihao\_lh@fudan.edu.cn}

\affil[1]{\orgdiv{Artificial Intelligence Innovation and Incubation Institute}, \orgname{Fudan University}, \orgaddress{\street{220 Handan Road}, \city{Shanghai}, \postcode{200433}, \state{Shanghai}, \country{China}}}

\abstract{
High-resolution wind information is essential for wind energy planning and power forecasting, particularly in regions with complex terrain. However, most AI-based weather forecasting models operate at kilometer-scale resolution, constrained by the reanalysis datasets they are trained on.
Here we introduce FuXi-CFD, an AI-based downscaling framework designed to generate detailed three-dimensional wind fields at 30-meter horizontal resolution, using only coarse-resolution atmospheric inputs. The model is trained on a large-scale dataset generated via computational fluid dynamics (CFD), encompassing a wide range of terrain types, surface roughness, and inflow conditions.
Remarkably, FuXi-CFD predicts full 3D wind structures—including vertical wind and turbulent kinetic energy—based solely on horizontal wind input at 10 meters above ground, the typical output of AI-based forecast systems. It achieves CFD-comparable accuracy while reducing inference time from hours to seconds.
By bridging the resolution gap between regional forecasts and site-specific wind dynamics, FuXi-CFD offers a scalable and operationally efficient solution to support the growing demands of renewable energy deployment.
}


\maketitle

\section{Introduction}\label{sec1}

The global energy transition is accelerating the deployment of renewable sources such as wind and solar. As the share of variable renewables increases, accurate and spatially resolved forecasts of wind conditions are becoming indispensable for ensuring system reliability, minimizing curtailment, and optimizing energy dispatch~\cite{wang2016quantifying}.
While offshore wind has attracted considerable attention, mountainous regions offer underutilized opportunities for wind power, owing to their natural wind acceleration effects, reduced land-use competition, and minimal environmental impact.
Harnessing this potential requires high-resolution wind predictions that can inform both turbine siting and operational decision-making~\cite{clifton2022research}.

Recent advances in artificial intelligence (AI) have led to a new generation of weather forecasting models—including GraphCast~\cite{lam2023learning}, FourCastNet~\cite{pathak2022fourcastnet}, Pangu-Weather~\cite{bi2023accurate}, FuXi~\cite{chen2023fuxi,chen2024machine}, MetNet~\cite{espeholt2022deep} and GenCast~\cite{price2023gencast}—that offer impressive improvements in forecast accuracy and lead time. These models are typically trained on reanalysis datasets such as ERA5~\cite{hersbach2020era5}, which provide a native spatial resolution of approximately 0.25° ($\sim$31 km) on a global grid. As a result, the outputs of AI weather models remain constrained to kilometer-scale resolutions. For wind energy applications, particularly in complex terrain, this resolution is insufficient: it fails to capture terrain-induced flow features and cannot support detailed power prediction or turbine-level control~\cite{dhunny2017wind}.

To obtain finer resolution, downscaling methods are often applied to coarse forecasts. Statistical approaches, based on interpolation, regression, or machine learning, provide computationally efficient corrections~\cite{schoof2013statistical} but struggle to generalize in regions lacking dense measurements and cannot reliably recover the three-dimensional, turbulent structures of wind flow~\cite{martinez2021review}.
Dynamical downscaling using mesoscale weather models such as WRF offers improved physical realism~\cite{soares2012wrf} but incurs high computational costs and suffers from limitations in turbulence parameterization and numerical stability in steep terrain~\cite{fernandez2018sensitivity}. These limitations have restricted their application in high-resolution operational wind forecasting, especially in mountainous regions~\cite{goger2016current}.

In contrast, computational fluid dynamics (CFD) models are widely used in wind energy engineering to simulate microscale wind behavior over complex terrain~\cite{porte2020wind}. CFD provides physically consistent predictions of wind speed and turbulence intensity at tens-of-meters resolution, making it a standard tool for wind farm feasibility studies (e.g., Meteodyn WT [1], WindSim [2], Envision Greenwich [3]). However, its computational expense and reliance on expert tuning make it impractical for routine forecasting or large-scale deployment.

To bridge the gap between high-fidelity physics and operational efficiency, we present FuXi-CFD, a deep learning–based downscaling framework trained on a large-scale dataset of CFD-generated wind fields. The model infers full three-dimensional wind and turbulence fields at 30-meter horizontal resolution and up to 300 meters in height, using only kilometer-scale wind components at 10 meters above ground level as input—matching the output format of modern AI weather models. By integrating CFD’s physical accuracy with the inference speed of AI, FuXi-CFD delivers terrain-aware, fine-scale predictions within seconds, providing a scalable solution for both wind farm planning and real-time power forecasting. It closes a long-standing resolution gap between AI-based weather models and the spatial demands of wind energy applications.

\begin{figure}[H]
\centering
\includegraphics[width=0.85\textwidth]{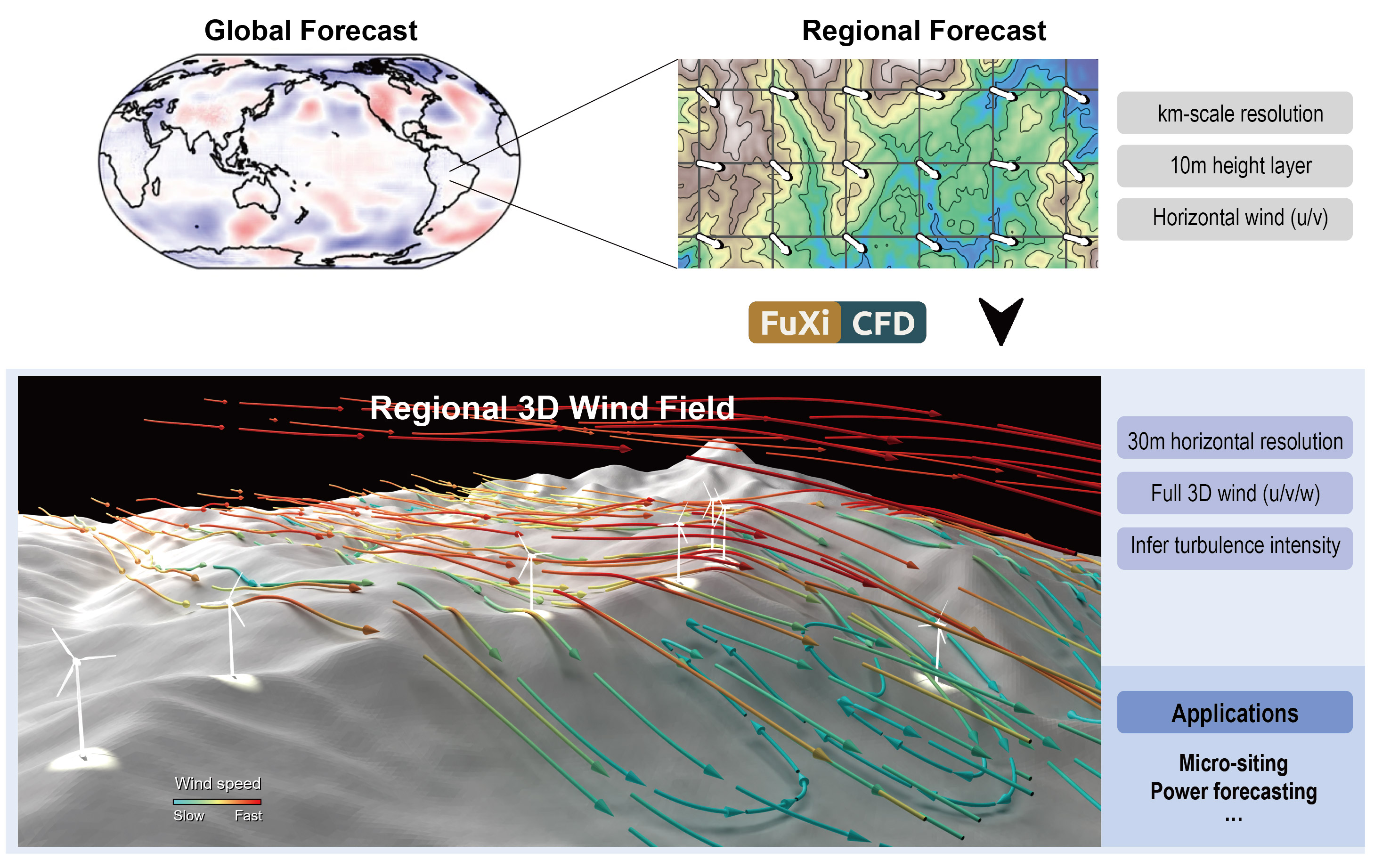}
\caption{
\textbf{Bridging the gap between AI-based regional weather forecasts and high-resolution wind field needs for wind energy.}
State-of-the-art AI weather models provide regional forecasts at kilometre-scale resolution and near-surface (10 m) horizontal wind components ($u$, $v$), which remain insufficient for wind energy applications requiring fine-scale three-dimensional wind information. By constructing terrain-aware CFD datasets, we enable the downscaling of regional forecasts to 30 m resolution full 3D wind fields ($u$, $v$, $w$), allowing for the inference of turbulence intensity and supporting applications such as micro-siting and power forecasting.}
\label{fig.graphicalabstract}
\end{figure}

\section{Methods}
\subsection{CFD-informed Dataset}

We construct the FuXi-CFD dataset as a high-fidelity resource to support the inverse reconstruction of wind fields over complex terrain. The dataset comprises over 12,000 steady-state CFD simulations, each representing a 9×9 km$^2$ region of real-world mountainous terrain. These scenarios span a wide range of surface roughness and elevation gradients that are representative of conditions relevant to wind energy siting and planning.

Each simulation is characterized by two physical inputs: elevation and surface roughness. Elevation is taken from the Shuttle Radar Topography Mission (SRTM) at 30-meter resolution~\cite{srtm1arcsecond}, while surface roughness is derived from land cover maps~\cite{zanaga2021worldcover} using empirical mappings to aerodynamic roughness length~\cite{hasager2003effective}. These two inputs are globally available and together define the terrain–atmosphere interaction context.

To generate diverse examples of wind–terrain interaction, we perform steady-state CFD simulations under various inflow conditions. Inflow velocity is initialized using a logarithmic atmospheric boundary layer profile with a 10 m/s reference wind at 100 m height at different directions. To support realistic flow development, each CFD domain is padded by 1.5 km in all directions beyond the central 9×9 km$^2$ region, yielding a full 12×12 km$^2$ simulation box. Simulations are performed using a Reynolds-Averaged Navier–Stokes (RANS) framework with standard turbulence models (see Supplementary Information for details).

Importantly, the FuXi-CFD dataset is not designed to replicate the forward process of CFD—that is, predicting wind fields from fully specified boundary conditions. Instead, it supports an inverse objective: generating terrain-sensitive 3D wind fields using only local terrain features and coarse-resolution meteorological data. To reflect this goal, the inflow wind speed and direction used during simulation are deliberately excluded from the dataset. This mirrors real-world wind energy planning scenarios, where upstream conditions are typically uncertain or unknown. In such cases, conventional CFD cannot be applied, but AI models trained on physically grounded simulations can still infer plausible wind patterns based on local context.

\begin{figure}[t]
\centering
\includegraphics[width=0.95\textwidth]{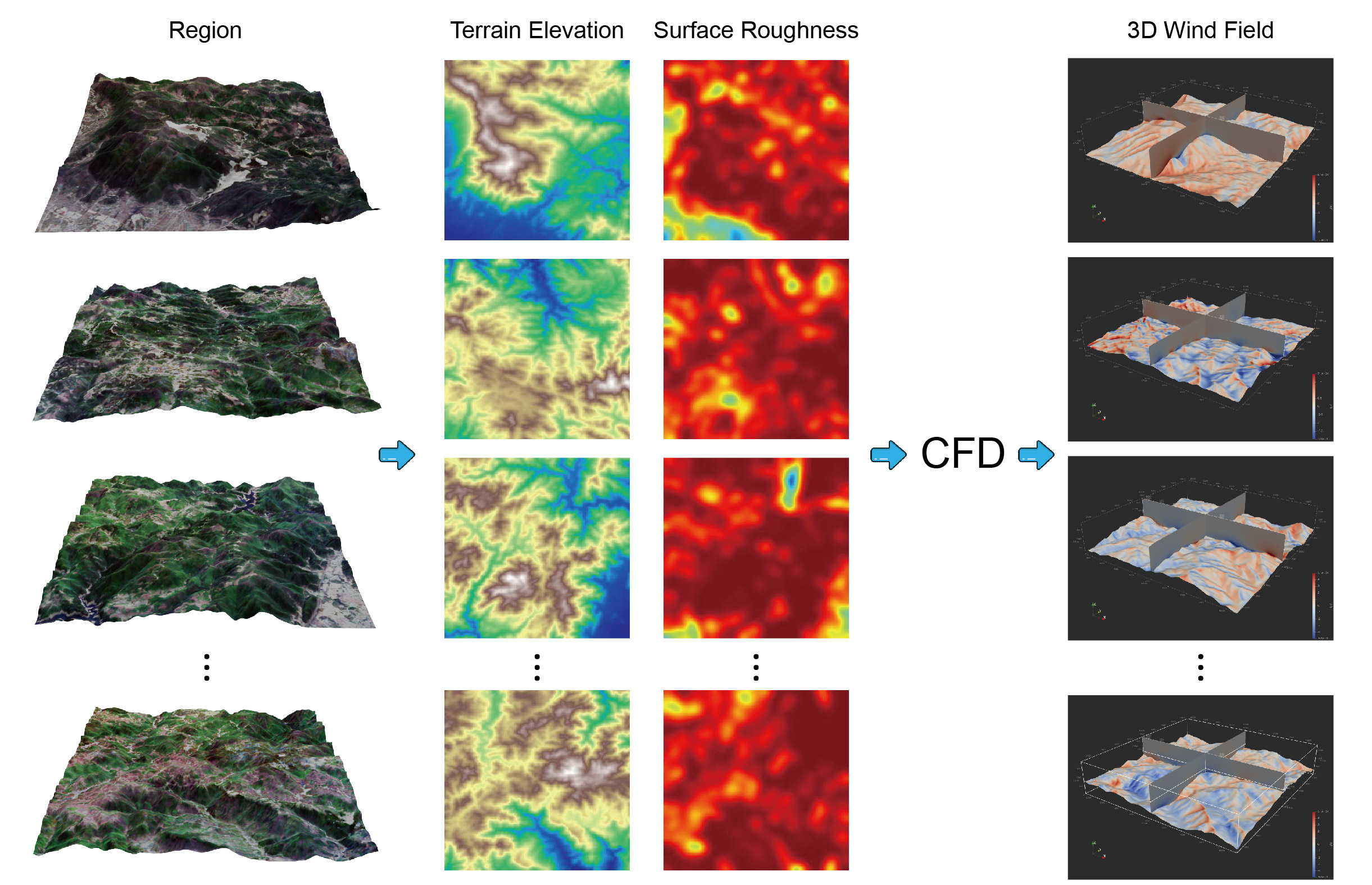}
\caption{
\textbf{Overview of the FuXi-CFD dataset generation workflow.}
Each row illustrates a representative mountainous region from the dataset. The first column shows a satellite view overlaid with elevation shading, providing an intuitive sense of terrain complexity. The second and third columns extract the two physical inputs used in simulation: terrain elevation from digital elevation models (DEMs) and surface roughness derived from land cover data. These inputs, along with a specified inflow direction, are used to drive CFD simulations of 3D wind fields over realistic terrain. The final column visualizes the simulated 3D wind field, shown as wind speed magnitude distributed over complex topography, together with two orthogonal vertical slices (in the xz and yz planes) to reveal internal flow structure.}
\label{fig.cfddataset}
\end{figure}

Figure~\ref{fig.cfddataset} shows several example cases from the dataset. The 3D wind fields exhibit complex spatial structures driven by terrain-induced effects such as acceleration, separation, and channeling. These patterns highlight the limitations of uniform downscaling approaches and underscore the need for terrain-aware modeling.

Each simulation outputs 3D fields of wind components ($u$,$v$,$w$) and turbulent kinetic energy ($k$) on a structured grid with 300×300×47 resolution up to 3 km altitude. Altogether, the dataset encompasses over 3 million CPU hours of computation and provides a large-scale, physically consistent training set for learning fine-scale wind dynamics.

\begin{figure}[t]
\centering
\includegraphics[width=0.9\textwidth]{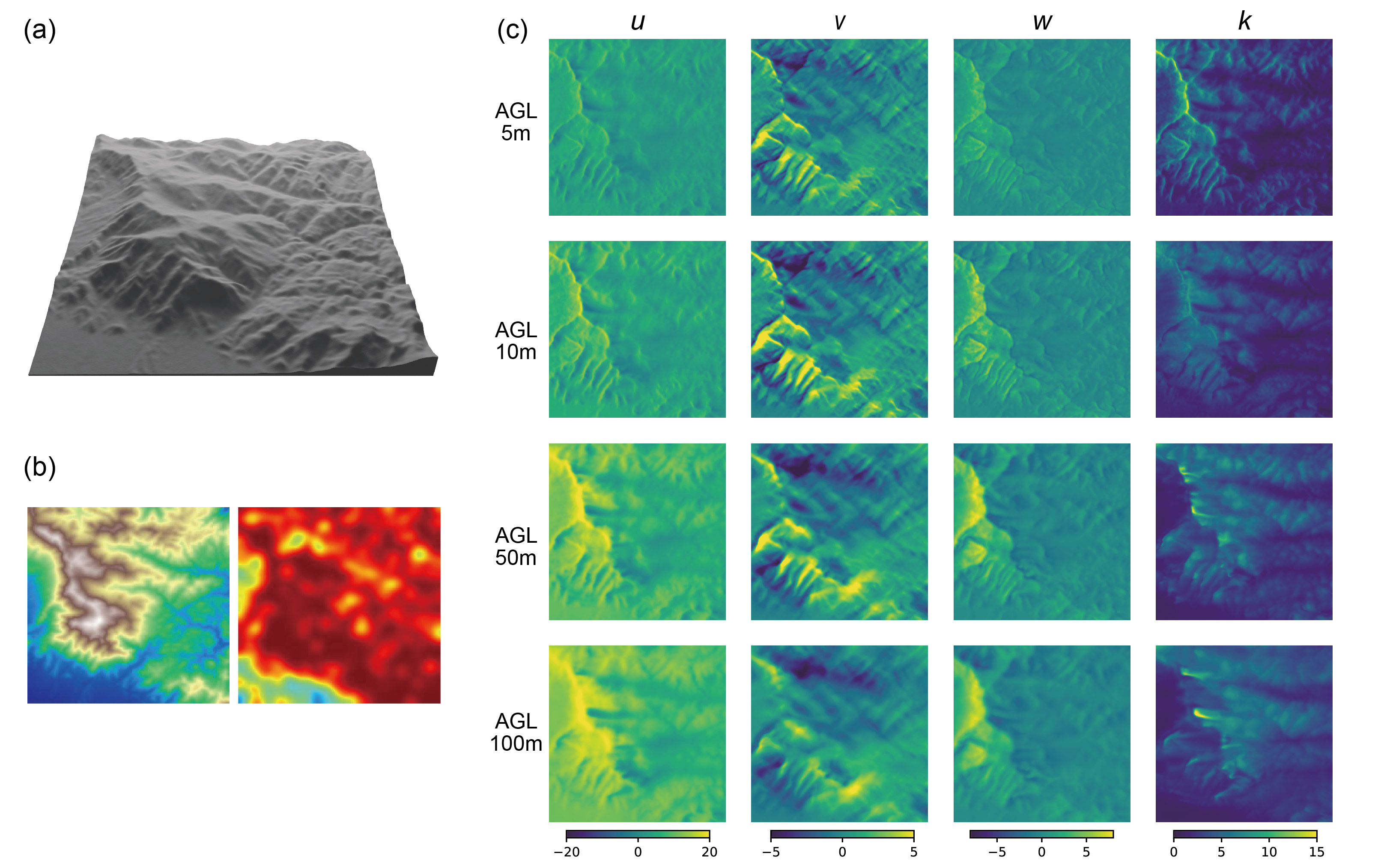}
\caption{
\textbf{CFD simulations reveal terrain-induced variability in wind and turbulence fields relevant for wind energy.}
(a) Terrain perspective view showing the region's topography. (b) Digital Elevation Model (DEM) and surface roughness maps used as inputs for the CFD simulations. (c) CFD results showing wind velocity components ($u$, $v$, $w$) and turbulent kinetic energy (k) at various heights above ground level (AGL). Higher altitudes generally correlate with higher wind velocities due to reduced terrain effects, consistent with atmospheric boundary layer (ABL) theory. However, at 100 m AGL, terrain effects remain noticeable, emphasizing the importance of terrain-aware modeling. Turbulent kinetic energy, crucial for wind energy applications, also shows significant spatial variation, with moderate turbulence being optimal for energy capture and excessive turbulence potentially reducing efficiency. The wind velocity components ($u$, $v$, $w$) and turbulent kinetic energy ($k$) exhibit complex, nonlinear patterns, driven by terrain-induced effects such as shear and separation zones.}
\label{fig.cfdexample}
\end{figure}

\subsection{AI-based downscaling model}

To reconstruct fine-scale three-dimensional wind fields from coarse atmospheric forecasts, we design a deep learning model that maps kilometer-scale near-surface wind inputs to high-resolution 3D wind structures.
Specifically, the model takes as input 10\,m AGL wind components ($u$, $v$) sampled on a $9 \times 9$ grid at 1\,km resolution, covering a $9 \times 9\,\mathrm{km}^2$ area. In parallel, terrain elevation and surface roughness maps are provided at 30\,m resolution over the same region.
These inputs represent the typical outputs of modern AI-based forecasting systems combined with satellite-derived surface data. The model is trained to infer four key physical variables—three wind components ($u$, $v$, $w$) and turbulent kinetic energy ($k$)—across 27 vertical levels up to approximately 300\,m, forming a four-channel volumetric output used for wind energy assessment.

\begin{figure}[t]
\centering
\includegraphics[width=0.99\textwidth]{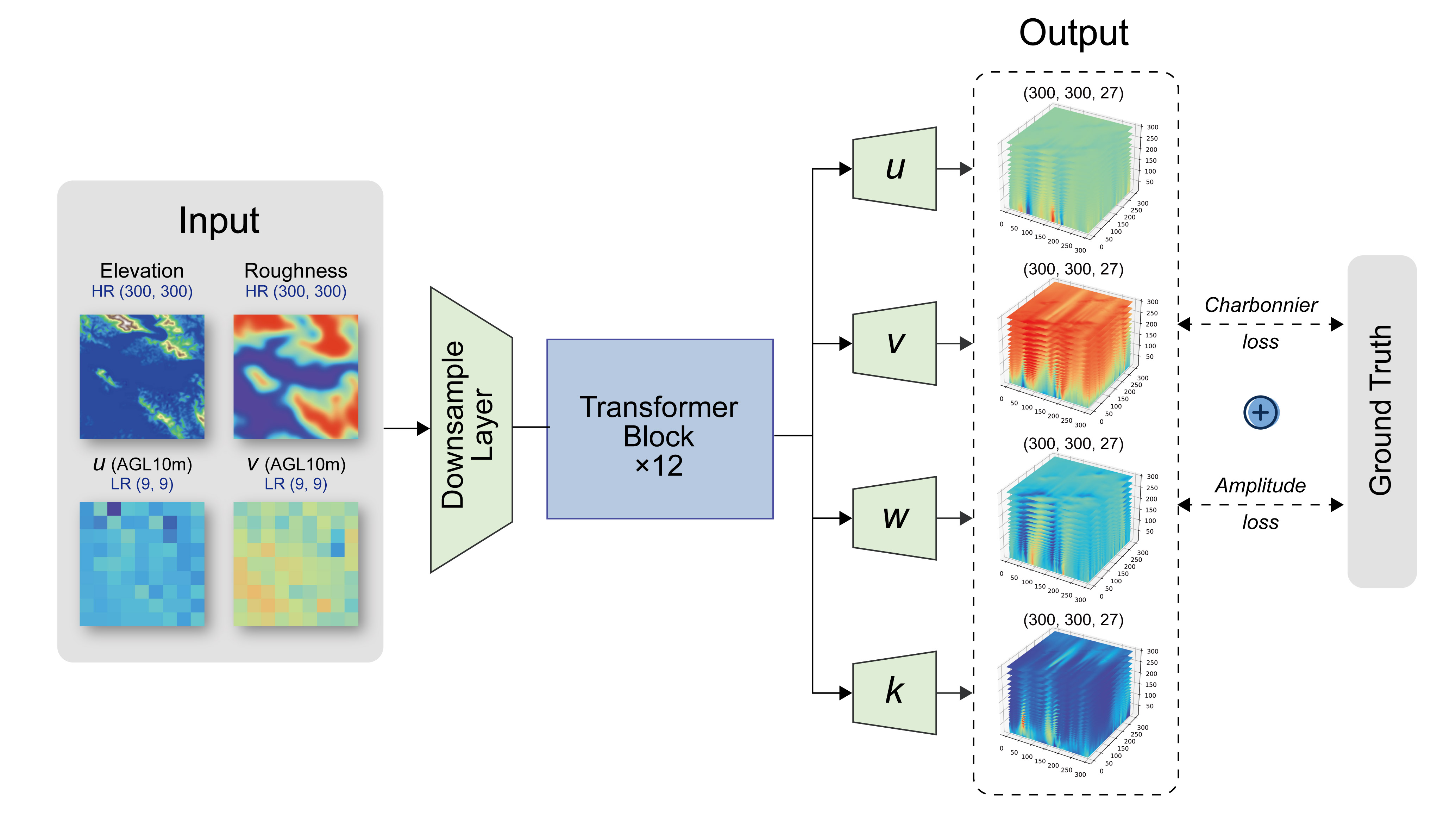}
\caption{
\textbf{Overall architecture of the terrain-aware deep learning model for 3D wind field downscaling.}
The model takes four inputs: low-resolution 10\,m wind components ($u$, $v$; 9$\times$9 grid), and high-resolution terrain elevation and surface roughness (300$\times$300 grid). A downsampling layer and a 12-block transformer encoder extract spatial features, which are decoded into three wind components ($u$, $v$, $w$) and turbulent kinetic energy ($k$) over 27 vertical levels. Each output has 30\,m horizontal resolution. The model is supervised by a combined Charbonnier and frequency-domain loss.
}
\label{fig.AI}
\end{figure}

We adopt a shared-encoder architecture with four task-specific decoder branches. The encoder is built upon a Vision Transformer backbone, which efficiently captures terrain-induced spatial correlations and long-range dependencies that are challenging for conventional convolutional models. The four input fields are concatenated into a tensor of shape (batch size, 4, 300, 300), corresponding to a 9$\times$9\,km$^2$ area at 30\,m resolution. The shared encoder extracts latent features from this input, which are then processed by separate decoder branches to reconstruct each target variable. An overview of the model structure is shown in Fig.~\ref{fig.AI}.

\begin{figure}[H]
\centering
\includegraphics[width=0.85\textwidth]{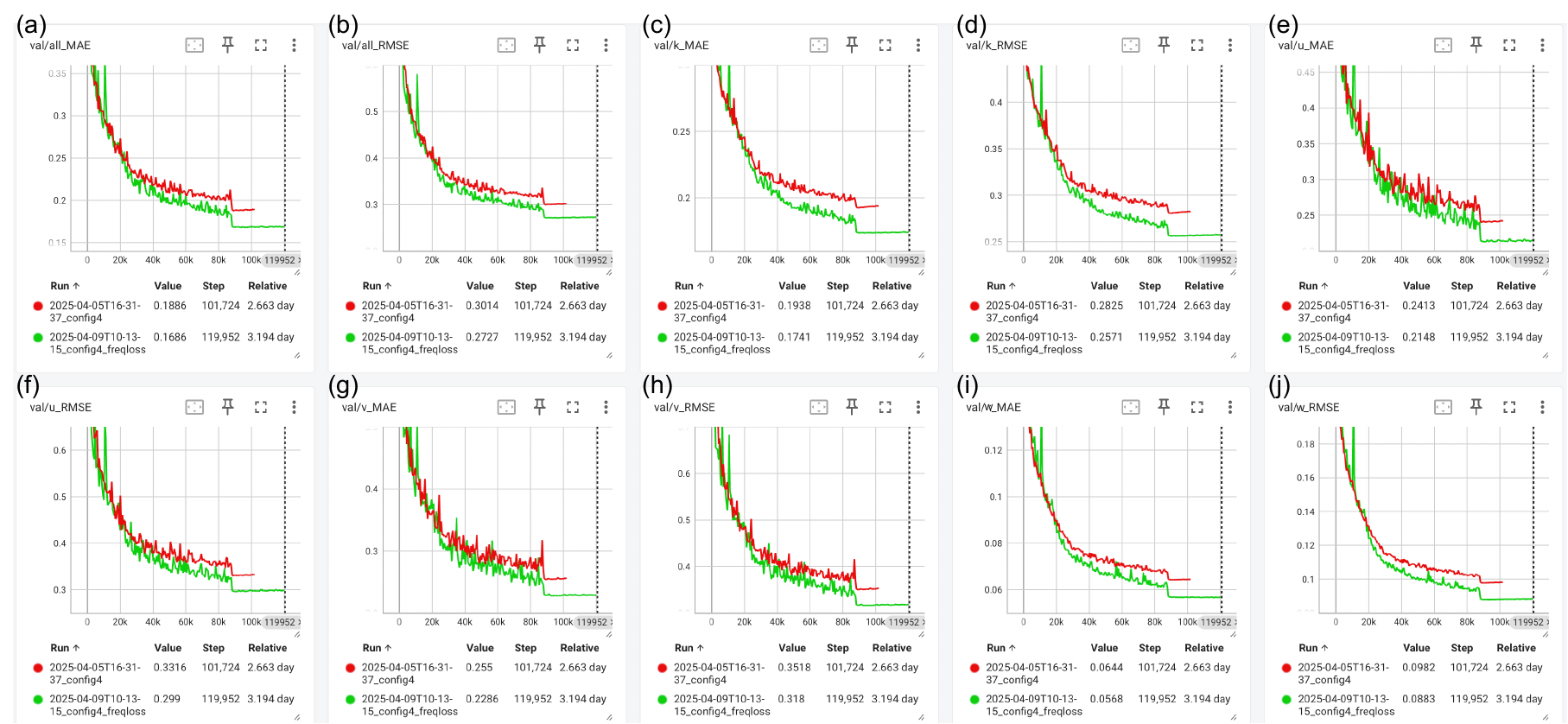}
\caption{
\textbf{Loss function design with combined spatial and frequency-domain supervision improves model performance.}
Validation metrics across training epochs for different output variables show that combining the Charbonnier loss with a frequency-domain loss (green lines) leads to consistently lower errors compared to using the Charbonnier loss alone (red lines). Subplots (a–j) show MAE and RMSE for total wind field, turbulent kinetic energy ($k$), and individual wind components ($u$, $v$, $w$). The joint loss leads to more stable convergence and improved accuracy across all targets.
}
\label{fig.loss}
\end{figure}

To enhance spatial fidelity and suppress noise in the predicted fields, we employ a hybrid loss function combining a robust spatial-domain loss with a frequency-domain regularization term. The former promotes local consistency, while the latter encourages the preservation of large-scale structural patterns. The model is trained using the AdamW optimizer with a stepwise learning rate schedule over approximately 128 hours on two A100 GPUs. As shown in Fig.~\ref{fig.loss}, this joint loss design leads to improved accuracy and more stable convergence across all target variables, particularly for the latent fields $w$ and $k$ that are not directly observable in the input.

Once trained, the model generates high-resolution 3D wind and turbulence fields within a fraction of a second per region, representing a speedup of over three orders of magnitude compared to traditional CFD simulations. This efficiency enables scalable deployment in wind energy applications such as turbine siting, power forecasting, and operational control.

\section{Results}

We evaluate FuXi-CFD's performance in predicting wind fields at multiple AGL heights, with a focus on spatial patterns, vertical profiles, and error characteristics.

\subsection{Prediction accuracy at key altitudes}

\begin{figure}[t]
\centering
\includegraphics[width=0.85\textwidth]{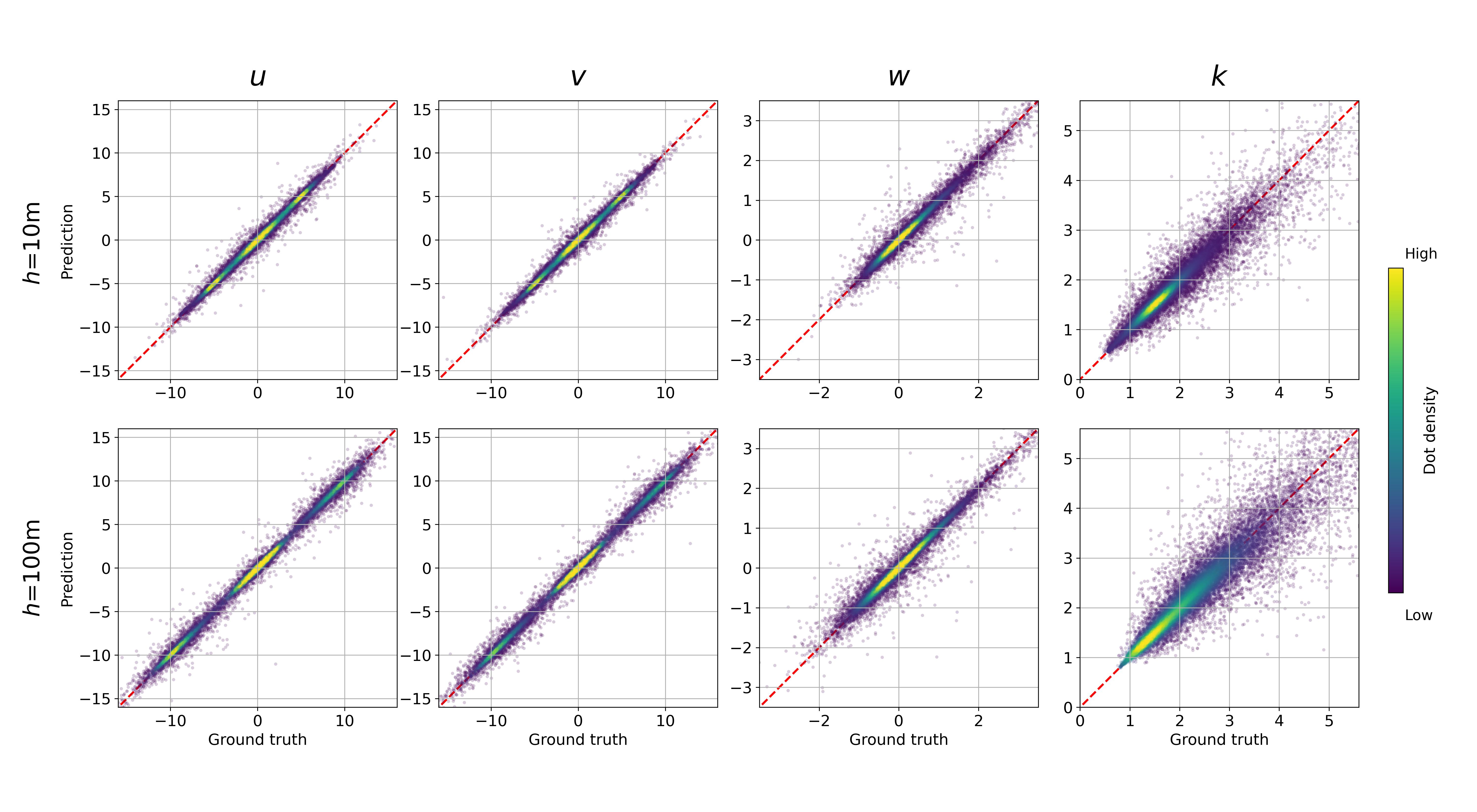}
\caption{
\textbf{Model predictions show high accuracy for wind components and reasonable performance for turbulence estimation.}
Parity plots compare model outputs with ground truth at 10 m and 100 m heights for horizontal wind components ($u$, $v$), vertical velocity ($w$), and turbulent kinetic energy ($k$). Dashed lines represent perfect agreement. Wind predictions align closely with ground truth, while predictions for $k$ show greater scatter, reflecting its nature as a derived, unobservable physical quantity.
}
\label{fig.parity}
\end{figure}

We first evaluated FuXi-CFD’s performance on unseen terrain by comparing its predictions with CFD results used as ground truth. The analysis focuses on four key physical quantities: horizontal wind speeds ($u,v$), vertical wind speed ($w$), and turbulent kinetic energy ($k$).
Figure~\ref{fig.parity} presents parity plots at two representative altitudes: 10\,m and 100\,m above ground level (AGL). These two layers were selected for their distinct significance—10\,m corresponds to the input height of $u$ and $v$, representing the level where the model receives direct wind information, while 100\,m is a critical height for wind energy applications, where wind speeds are typically higher and terrain-induced effects are more complex.
The red diagonal line in each plot indicates perfect prediction.

As shown in the Fig.~\ref{fig.parity}, FuXi-CFD achieves high accuracy for horizontal wind components ($u,v$), especially in the mid-to-high speed range where the predictions closely align with the ground truth and exhibit limited scatter. Performance for the vertical component ($w$) is slightly less accurate but remains reasonable. Predictions for turbulent kinetic energy ($k$) show the largest deviations, with noticeable dispersion.

This lower accuracy in predicting $k$ can be attributed to the fact that, while $k$ is included in the training labels, it is not provided as an explicit input to the model. As a result, $k$ functions as an inferred or latent variable that must be deduced from other input features—making it inherently more difficult to predict.

A comparison between the 10\,m and 100\,m results also reveals a meaningful physical trend: while all wind components ($u,v,w$) increase significantly with height, the turbulent kinetic energy $k$ tends to decrease. This aligns with practical understanding in wind energy siting, where turbines are typically installed at heights with stronger and more stable winds, characterized by lower turbulence levels.

\subsection{Wind pattern comparison across heights}

\begin{figure}[H]
\centering
\includegraphics[width=0.82\textwidth]{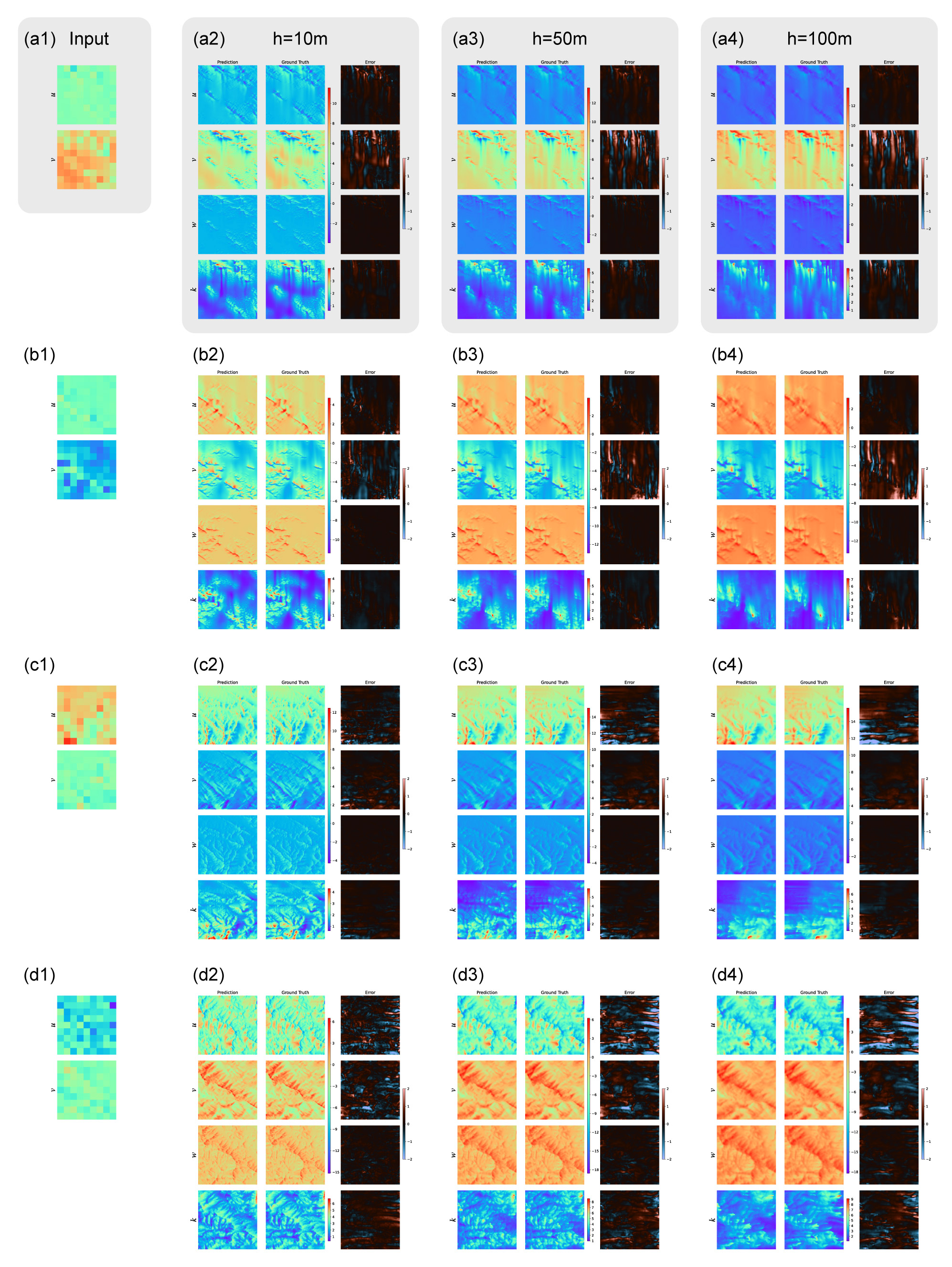}
\caption{
\textbf{Multi-height wind field reconstructions across diverse terrain conditions.}
Four representative cases illustrate model performance under diverse terrain and inflow conditions. Each example shows the coarse input wind field (9\,×\,9 grid at 10\,m AGL), followed by high-resolution predictions and CFD references at 10\,m, 50\,m, and 100\,m AGL. Wind speed increases and flow smoothness improves with altitude, consistent with reduced terrain influence. Despite a downscaling ratio exceeding 30×, the model captures fine-scale wind structures with high fidelity. Beyond horizontal components ($u$, $v$), FuXi-CFD also infers vertical velocity ($w$) and turbulent kinetic energy ($k$), both critical but absent from the input.
}
\label{fig.pattern}
\end{figure}

We examine the spatial patterns of FuXi-CFD predictions at multiple altitudes to assess how terrain-induced effects are captured across different physical quantities. Figure~\ref{fig.pattern} presents four representative test cases with varying terrain types and inflow conditions. For each case, the first column (a1, b1, c1, d1) shows the low-resolution input wind field (9×9 km at 10\,m AGL). The remaining columns (a2–a4, b2–b4, etc.) display high-resolution predictions at 10\,m, 50\,m, and 100\,m AGL, alongside corresponding CFD references and absolute error maps.

As altitude increases, wind speeds generally rise and flow patterns become smoother, consistent with reduced terrain influence in the atmospheric boundary layer. Across all cases, FuXi-CFD predictions align closely with CFD references, maintaining the overall spatial structure and local variability of the wind field. Errors remain low and show no systematic over- or under-estimation, indicating that the model captures both the magnitude and spatial pattern without introducing bias. Relative errors are particularly low at near-surface levels and increase slightly with altitude, where finer-scale terrain effects become less dominant.

Importantly, in addition to downscaling horizontal wind components ($u$, $v$), FuXi-CFD also infers vertical velocity ($w$) and turbulent kinetic energy ($k$)—both entirely absent from the input. The accurate recovery of these quantities further demonstrates the model’s ability to generalize to unobserved physical variables and reconstruct coherent three-dimensional wind fields.

\subsection{Error analysis at different AGL heights}

\begin{figure}[b]
\centering
\includegraphics[width=0.8\textwidth]{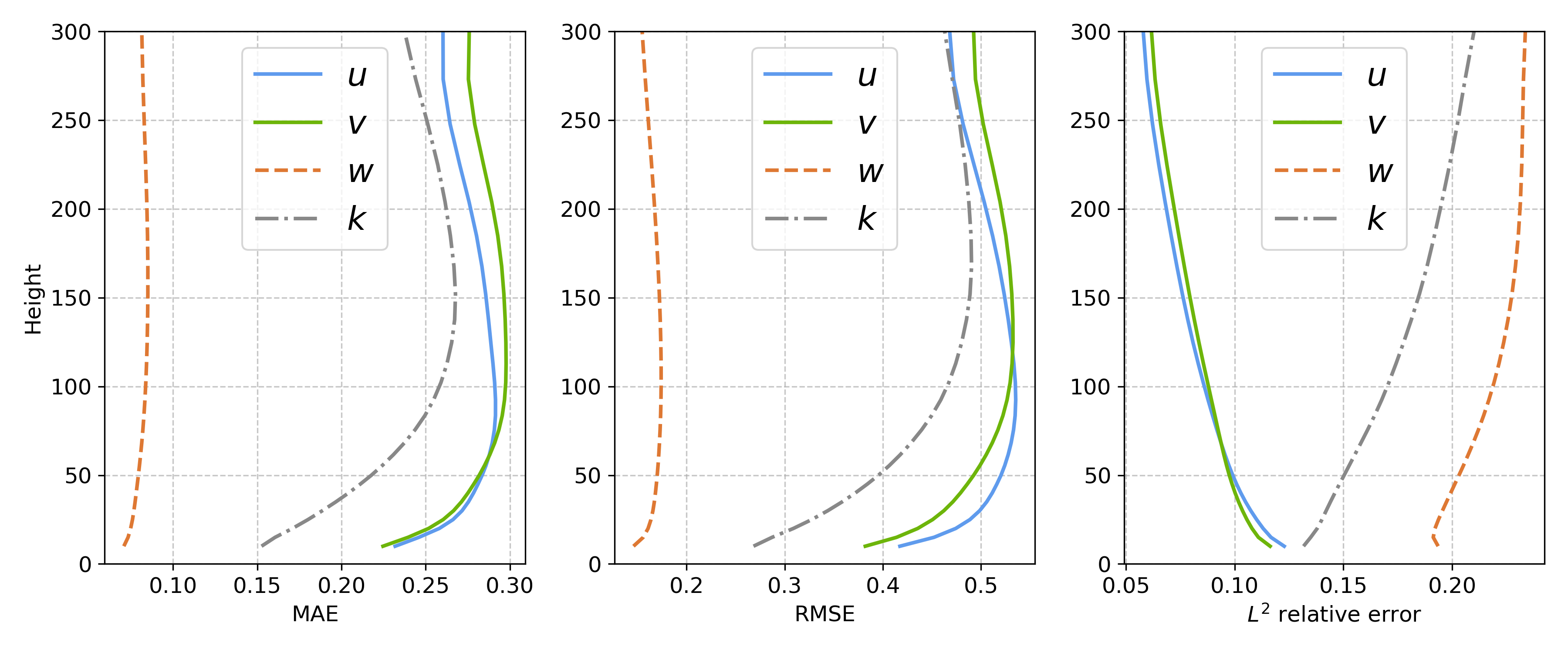}
\caption{
\textbf{Error analysis of FuXi-CFD predictions across altitudes from 0 to 300\,m above ground level (AGL).}
(a) Mean absolute error (MAE) and (b) root mean square error (RMSE) are evaluated for four flow variables: horizontal wind components u and v, vertical velocity w, and turbulent kinetic energy k. Prediction errors are lowest near the surface, where high-resolution terrain elevation and surface roughness serve as strong constraints in the model inputs. Errors increase around 100\,m, a transition zone where the influence of terrain weakens while flow complexity remains high, resulting in reduced model accuracy. At higher altitudes, errors decrease again due to smoother flow structures and reduced spatial variability. Notably, the 100\,m level—critical for wind energy applications—shows the highest error, highlighting the challenge of capturing mid-layer flow dynamics.}
\label{fig.error.profile}
\end{figure}

Further, we quantitatively analyze the statistical error behavior of FuXi-CFD across different AGL levels.
Figure~\ref{fig.error.profile} reports the MAE, RMSE, and L2 relative errors of key wind field components from 5\,m to 300\,m.

Both MAE and RMSE peak near 100\,m, while lower and higher altitudes show reduced errors. Near the surface, this is largely due to lower wind speeds yielding smaller absolute errors. At higher altitudes, terrain-induced complexity weakens, and the flow becomes smoother and more predictable. The elevated error near 100\,m reflects the challenge of modeling in the transition layer where both surface effects and flow complexity coexist—precisely the region most critical to wind energy applications.

The close alignment of MAE and RMSE profiles suggests that error distributions remain consistent with height, without disproportionate contributions from large outliers at any particular level. The near-constant RMSE-to-MAE ratio further supports the statistical stability of model performance across the vertical domain.

L2 relative errors, in contrast, exhibit a monotonic decline with height, as increasing wind speeds in the upper ABL reduce relative deviations. At 100\,m, relative errors in horizontal wind components fall below $7\%$, while vertical components ($w$ and $k$) reach $~20\%$. Given the complexity of mountainous flows, and the fact that w and k are latent quantities inferred without direct input, this level of accuracy represents a meaningful advance over existing downscaling approaches, especially given the lack of direct input for latent quantities like $w$ and $k$.

\subsection{Point-wise wind profile comparison}

\begin{figure}[H]
\centering
\includegraphics[width=0.99\textwidth]{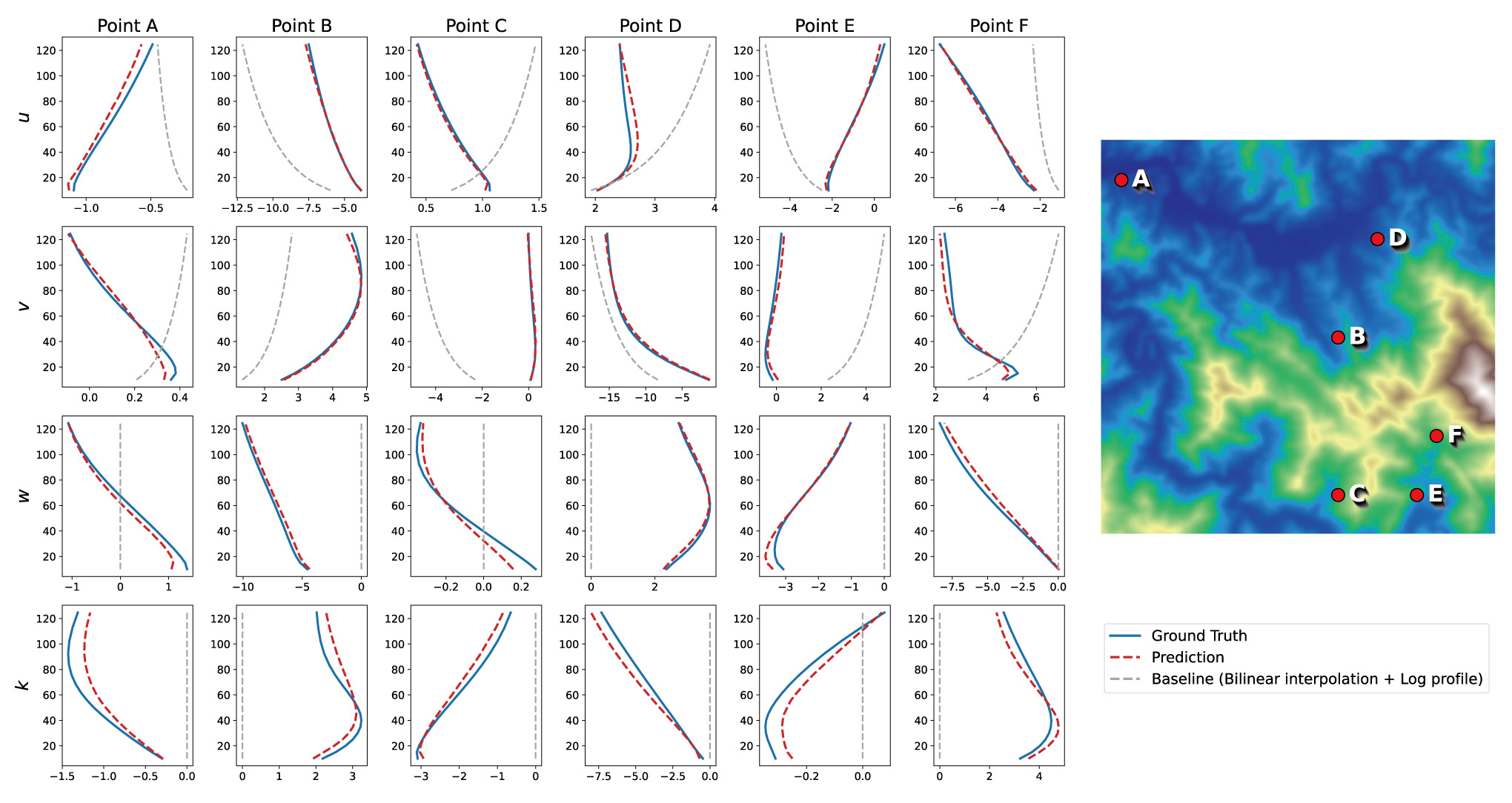}
\caption{
\textbf{Vertical wind profiles at representative locations.}
Predicted and CFD reference vertical profiles of horizontal wind components ($u$, $v$), vertical velocity ($w$), and turbulent kinetic energy ($k$) are shown at six locations, indicated by red circles on the map. The results demonstrate that the model captures the general trends and magnitudes of the CFD data across all variables.
}
\label{fig.points.profile}
\end{figure}

We further examine FuXi-CFD’s ability to reconstruct vertical wind profiles at specific locations.
Figure~\ref{fig.points.profile} presents representative results from six randomly selected points within a mountainous region, with background elevation indicated on the map. For each location, the vertical profiles of the four velocity-related quantities ($u$, $v$, $w$, and $k$) are plotted. The FuXi-CFD predictions (red dashed lines) are compared against high-fidelity CFD simulations (blue solid lines). As a baseline, u and v profiles derived from bilinear interpolation of the 10\,m wind input, followed by logarithmic extrapolation, are also shown (grey lines). No baseline is available for $w$ and $k$ due to the absence of direct inputs.

The results show that FuXi-CFD closely tracks the CFD reference across all components, effectively capturing flow trends and site-specific features. In contrast, the baseline method fails to represent local structure or vertical variation. Particularly notable is the model’s ability to infer latent variables ($w$ and $k$) without upstream input. Given the absence of state-of-the-art methods capable of downscaling kilometer-scale data to ten-meter resolution with such fidelity, this represents a substantial advance in data-driven atmospheric modeling.

\section{Conclusion and Discussion}

This study presents FuXi-CFD, an AI-based downscaling model that enables three key capabilities not simultaneously achieved in prior work. First, it captures terrain-induced effects on wind fields, a critical factor in realistic wind resource assessment. Second, it downscales kilometer-scale atmospheric inputs to 30-meter resolution 3D wind fields, allowing the recovery of fine-scale flow structures that are essential for site-specific applications. Third, it delivers this level of resolution at a speed more than three orders of magnitude faster than conventional CFD simulations, making high-resolution wind predictions computationally practical for large-scale and real-time deployment.

Beyond these technical contributions, FuXi-CFD addresses a central challenge in renewable energy forecasting: the scale mismatch between weather models and the resolution requirements of wind energy projects. By bridging this gap, FuXi-CFD provides a scalable solution for integrating high-resolution wind field predictions into operational workflows such as wind farm siting, power forecasting, and grid integration. The model complements emerging AI-based weather systems by serving as a lightweight, terrain-aware post-processor that adds critical spatial detail without significant computational burden.

While the current model is trained on steady-state simulations under neutral stability conditions, future extensions to include atmospheric stratification and unsteady flows would enhance its generalizability. Nevertheless, this work demonstrates a viable path toward high-fidelity, energy-efficient wind forecasting at operational scales—supporting the broader adoption of wind power in diverse and complex environments.

\backmatter

\bibliography{sn-bibliography}

\end{document}